\documentclass[aps,prl,twocolumn]{revtex4-2}
\usepackage{latexsym}
\usepackage{balance}
\usepackage{amsmath}

\usepackage{amssymb}
\usepackage{bm}
\usepackage{wasysym}
\usepackage[dvips]{color}
\usepackage{graphicx}
\usepackage{subfigure}
\usepackage{epsfig}
\usepackage{balance}
\usepackage{subfigure}
\usepackage{ulem}
\usepackage[LGR,T1]{fontenc}
\usepackage{bm}

\begin{document}

\title{Aharonov-Casher phase in twisted bilayer graphene}

\author{Igor Kuzmenko$^{1,2}$, Y. B. Band$^{1,2,3}$, Yshai Avishai$^{1,4}$}

\affiliation{
  $^1$Department of Physics,
  Ben-Gurion University of the Negev,
  Beer-Sheva 84105, Israel
  \\
  $^2$Department of Chemistry,
  Ben-Gurion University of the Negev,
  Beer-Sheva 84105, Israel
  \\
  $^3$The Ilse Katz Center for Nano-Science,
  Ben-Gurion University of the Negev,
  Beer-Sheva 84105, Israel
  \\
  $^4$Yukawa Institute for Theoretical Physics, Kyoto, Japan
  }
  
\begin{abstract}
The Aharonov-Casher (AC) effect is a quantum mechanical phenomenon in which the wave function of a  particle with a magnetic moment moving  in a region subject to  an electric field develops a phase shift due to spin-orbit interaction, even if no classical force acts on it. This phase also depends on the medium through which the particle moves. Here we focus on the AC phase of an electron moving in twisted bilayer graphene (TBG) lying in the $x$-$y$ plane, subject to a uniform electric field perpendicular to the plane of the graphene, ${\bf E}=E{\hat{\bf z}}$. The AC phase is determined by an $SU(2)$ vector potential ${\bf A}$ from which a phase factor is generated, and used to perform a gauge transformation of the Hamiltonian.  We find that the AC phase for a straight line path between two points in the TBG plane is linear with $E$ and exhibits sharp peaks at the magic angles.  To help demonstrate an experimental method for determining the AC phase, we examine the probability of polarized electron propagation from a source tip to a drain tip in a double-tip scanning tunneling spectroscopy configuration.
\end{abstract}

\maketitle

{\it Introduction}---The state of electrons in single-layer and bilayer graphene with momenta close to the Dirac point is special. This is where the material's conduction and valence bands meet, causing electrons and holes to behave as massless Dirac fermions. In this region, charge carriers are highly sensitive to disorder, leading to the formation of electron-hole ``puddles'' and unusual quantum transport phenomena. This unique behavior makes the momentum region near the Dirac point ideal for studying relativistic quantum mechanics and for potential electronic applications \cite{Kane-Mele-2005, Kane-Mele-QSHE-2005}.  Dirac massless fermions are found in many condensed matter systems beyond graphene, including group IV monolayers such as silicene and germanene \cite{Kamal_19}, the surfaces of 3D topological insulators \cite{Crepel_21}, and specific quantum well systems, such as HgTe/CdTe \cite{Krishtopenko_19}, among others. These systems are similar to graphene in the sense that they exhibit low-energy linear dispersion relations and characteristic behavior such as Klein tunneling. This offers unique opportunities to study Dirac physics in various materials. 

Near the Dirac point, twisted bilayer graphene (TBG) exhibits unique phenomena such as nearly vanishing Fermi velocity $v_F(\theta)$ and nearly flat Brillouin bands \cite{Wei_25}, as well as the emergence of correlated phases, such as superconductivity and magnetism, at magic angles \cite{Cao_18, Roy_19}. Unlike single-layer graphene, TBG develops flat electronic bands at these specific angles, which leads to strong electron interactions that can drive these exotic phases. The presence of disorder and in-plane magnetic fields introduces further complexity because these factors can modify the Dirac points and lead to other interesting behaviors. 

The Aharonov-Casher (AC) effect \cite{AC_84}, proposed by Aharonov and Casher in 1984, is a quantum phenomenon that clarifies the role of geometric phases in quantum mechanics and demonstrates how the presence of an electric field affects the quantum states of particles with a magnetic moment.  For a particle with a magnetic moment moving in an external electric field, the AC effect shows that the particle's wave function acquires a geometric phase that depends on the path taken by the particle \cite{Kuzmenko_25}.  This leads to observable consequences in interference patterns.  The AC effect has applications in various fields of quantum mechanics, especially interferometry. For instance, it can manipulate and control the phase of particle waves in interferometers. This provides a mechanism for precisely measuring physical quantities, such as the magnetic dipole moment of atoms and particles \cite{Cimmino_89}. Advancements in experimental techniques have enabled high-precision observation of the AC effect \cite{Gillot_14}.

The AC phase in graphene has been investigated theoretically \cite{Recher_07}, and experimentally \cite{Russo_08} primarily in the context of graphene-based nano-rings where it is induced by the Rashba spin-orbit interaction (SOI) \cite{Ghaderzadeh_18}. The phase manifests as measurable oscillations in the non-equilibrium current through the nano-ring devices, demonstrating the effect of electromagnetic potentials on electron transport in graphene.

In this Letter we explore the behavior of the AC phase $\varphi_{\mathrm{AC}}$ as function of the electric field strength $E$ and twist angle $\theta$, and show that $\varphi_{\mathrm{AC}}$ exhibits sharp peaks at the magic angles, with peak height inversely proportional to $v_F(\theta)$, and grows linearly with $E$, and propose a double-tip scanning tunneling spectroscopy method of measuring the AC phase.

{\it Twisted Bilayer Graphene}---We assume that the two graphene layers, located on the $x$-$y$ plane, are rotated by an angle $\pm \theta/2$ about the $z$ axis, respectively, and are subject to a uniform electric field ${\bf E}=E{\hat{\bf z}}$.  Denote the Pauli matrices for sub-lattice A/B in a given layer by ${\bm \tau}=(\tau_x,\tau_y,\tau_z)$ , the Pauli matrices for the two layers by ${\bm \eta}=(\eta_x,\eta_y,\eta_z)$ , and the three Pauli matrices for the electron spin by ${\bm \sigma}$.  The Hamiltonian for an electron in the TBG can be written as an $8$$\times$$8$ matrix ${\bm \tau} \otimes {\bm \eta} \otimes {\bm \sigma}$ (A/B~sub-lattices$\otimes$two-layers$\otimes$spin-space) \cite{Song-21},
\begin{equation}   \label{eq:H=H_0+H_M+H_R}
H(\mathbf{r}) = H_0 + H_m(\mathbf{r}) + H_R .
\end{equation}
Here the first term is the kinetic energy Hamiltonian of the TBG.  For small twist angles ($\theta \ll 1$~rad), $H_0$ takes the form
\begin{align}
H_0 &=
v_0
\big(
  \eta_0 \, \sigma_0 \, {\bm p} \cdot {\bm \tau} -
  \frac{\theta}{2} \, \eta_z \, \sigma_0 \, [ {\bm p} \times {\boldsymbol \tau} ]\cdot \hat{\bm{z}} 
\big) \nonumber \\
&= v_0
\begin{pmatrix}
  h({\bm p}_{\theta/2}) & 0 \\ 0 & h({\bm p}_{-\theta/2})
\end{pmatrix} ,
\end{align}
where ${\bf p} =-i \hbar {\bm \nabla}$, $\bm{p}_{\pm \theta/2} = \cos(\tfrac{\theta}{2}) \, \bm{p} \mp  \sin(\tfrac{\theta}{2})\, \hat{\bm{z}} \times \bm{p} \approx \bm{p} \mp \frac{\theta}{2} \, \hat{\bm{z}} \times \bm{p}$ is the momentum vector rotated through the angle $\pm \theta/2$ around the $z$-axis,
$\eta_0 = \begin{pmatrix} 1 & 0 \\ 0 & 1 \end{pmatrix}$, $\eta_z = \begin{pmatrix} 1 & 0 \\ 0 & -1 \end{pmatrix}$, and the 4$\times$4 matrices $h({\bm p}_{\pm \theta/2})$ are the electron kinetic energy in the first ($\theta/2$) or second ($-\theta/2$) graphene layer,
\begin{equation}   \label{eq:h_single_layer}
h({\bm p}_{\pm \theta/2}) =
\sigma_0 \, {\bm p}_{\pm \theta/2} \cdot {\bm \tau} =
\sigma_0 \, {\bm p} \cdot {\bm \tau}  \mp
\frac{\theta}{2} \, \sigma_0 \, [ {\bm p} \times {\boldsymbol \tau} ] \cdot \hat{\bm z} .
\end{equation}

The second term on the right hand side of Eq.~(\ref{eq:H=H_0+H_M+H_R}), $H_m(\mathbf{r})$, is the the moir\'e potential Hamiltonian,
\begin{equation}
H_m(\mathbf{r}) =
\sigma_0 \big[ T(\mathbf{r}) \eta^+ + T^{\dagger}(\mathbf{r}) \eta^- \big] =
\sigma_0 \begin{pmatrix} 0 & T({\bf r}) \\ T^\dagger({\bf r}) & 0 \end{pmatrix} .
\end{equation}
Here $\eta^{\pm} = \tfrac{1}{2} ( \eta_x \pm i \eta_y)$, and the moir\'e potential is given by
\begin{equation}   \label{eq:T-def}
T({\bf r})=\sum_{j = 1}^3 e^{-i{\bf q}_j \cdot {\bf r}}T_j .
\end{equation}
Here
\begin{equation}   \label{eq:q_1-q_2-q_3}
  {\bf q}_j = k_\theta
  \Big(
    \sin \Big[ \frac{2 \pi (j - 1)}{3} \Big] \, \hat{\mathbf{x}} -
    \cos \Big[ \frac{2 \pi (j - 1)}{3} \Big] \, \hat{\mathbf{y}}
  \Big) ,
\end{equation}
where $k_\theta$ is the distance between $K$ and $K'$ points in the first moir\'e Brillouin zone,
\begin{equation}
  \label{eq:k_theta}
  k_\theta = 2 k_D \, \sin\tfrac{\theta}{2} ,
  \quad
  k_D = \frac{4\pi}{3a} = 1.70276 \, \text{\AA}^{-1} .
\end{equation}
The 4$\times$4 matrices $T_j$ are
\begin{equation}   \label{eq:T_j}
  T_j = w_0 \tau_0 +
 w_1 \Big[ \tau_x \cos \Big( \frac{2 \pi(j-1)}{3} \Big) + \tau_y \sin \Big( \frac{2 \pi(j-1)}{3} \Big) \Big] ,
\end{equation}
where $w_0=77.0371 \, {\rm meV}$ and $w_1=110.053 \, {\rm meV}$ are the two coupling constants appearing in Eq.~(\ref{eq:T_j}), $\hbar v_0=659.107 \, {\rm meV \, nm}$ and $a = 2.46 \, \text{\AA}$ is the graphene lattice constant. Since the $w_0$ term contributes to the diagonal elements, it represents the interlayer coupling between the A/B sub-lattices of layers 1 and 2 \cite{Song-21}.  Similarly, the $w_1$ term only contributes to the off-diagonal elements; it is thus associated to the interlayer coupling between the A/B sub-lattices of layers 1 and 2 \cite{Song-21}.

The third term on the right hand side of Eq.~(\ref{eq:H=H_0+H_M+H_R}), $H_R$, is the Rashba interaction Hamiltonian,
\begin{align}
H_R &=
- \lambda_R
\big(
  \eta_0 \, [\hat {\bf z} \times {\bm \sigma}] \cdot {\boldsymbol \tau} -
  \frac{\theta}{2} \, \eta_z [[\hat {\bf z} \times {\bm \sigma}] \times {\boldsymbol \tau}]_z
\big) \nonumber \\
&=
- \lambda_R
\begin{pmatrix}
  [\hat {\bf z} \times {\bm \sigma}_{\theta/2}] \cdot {\boldsymbol \tau} &
  0
  \\
  0 &
  [\hat {\bf z} \times {\bm \sigma}_{-\theta/2}] \cdot {\boldsymbol \tau}
\end{pmatrix} ,
\end{align}
where ${\bm \sigma}_{\pm \theta/2} \approx {\bm \sigma} \mp \frac{\theta}{2} \, \hat{\bm{z}} \times {\bm \sigma}$,
$\lambda_R = \tfrac{\xi e  z_0}{3 V_{s, p,\sigma}} E$, is the Rashba coupling strength, which has units of energy, $e$ is the elementary electric charge, $z_0 \approx 3.5 a_B$ is the matrix element defined by the spatial overlap of the carbon atomic orbitals, $a_B$ is the Bohr radius, $V_{s,p,\sigma} = 5.580$ eV is the tight-binding hopping parameter for graphene, $\xi \approx 6$ meV is the intrinsic atomic spin-orbit coupling strength of an isolated carbon atom \cite{Min_11}.

{\it Low-energy effective Hamiltonian and Fermi velocity---}References~\cite{lopes2007graphene, bistritzer2011moire, Tarnopolsky_19} developed a continuum model to derive low-energy effective Hamiltonian for TBG with a `renormalized' Fermi velocity $v_F(\theta)$ (the moir\'e potential gives rise to a Fermi velocity that depends on $\theta$ and is referred to as renormalized in Ref.~\cite{bistritzer2011moire}).
The effective Hamiltonian including the Rashba interaction term is
\begin{equation}   \label{eq:h_eff-def}
h_{\text{eff}}(\hbar \mathbf{k}) =
\hbar v_F(\theta) \, \sigma_0 \, \bm{\tau} \cdot \bm{k}
- \lambda_R \, {\boldsymbol \tau} \cdot  [\hat {\bf z} \times {\bm \sigma}] ,
\end{equation}
where the renormalized Fermi velocity was derived in Refs.~\cite{lopes2007graphene, bistritzer2011moire, Tarnopolsky_19} using different continuous models.

The electronic structure of TBG has been modeled for small twist angles by Bistritzer and MacDonald~\cite{bistritzer2011moire}. In their continuum model, the interaction between the layers causes a hybridization of the Dirac cones that are shifted in momentum space by a distance $k_\theta$, which has been defined in Eq.~(\ref{eq:k_theta}). The Bistritzer-MacDonald (BMD) continuum model took $w_0 = w_1 \equiv w \approx 110$~meV, and applied perturbation theory to the effective Hamiltonian near the moir\'{e} Dirac points obtaining the following analytical formula for the ratio of the renormalized Fermi velocity $v_F(\theta)$ to the monolayer velocity $v_0$,
\begin{equation}   \label{eq:fermi_velocity-Bistritzer}
\frac{v_F(\theta)}{v_0} = \frac{1 - 3 \alpha^2(\theta)}{1 + 6 \alpha^2(\theta)} ,
\end{equation}
where, $\alpha(\theta)$ is a dimensionless parameter characterizing the effective interlayer coupling strength,
\begin{equation}
\alpha(\theta) = \frac{w}{\hbar v_0 k_\theta} = \frac{w}{2\hbar v_0 k_D \sin(\theta/2)} .
\end{equation}
The Fermi velocity in Eq.~(\ref{eq:fermi_velocity-Bistritzer}) vanishes at the first magic angle, which is equal to $\theta_1 = 1.05^\circ$ \cite{bistritzer2011moire}.  Reference~\cite{bistritzer2011moire} also calculated other magic angles, $\theta_2 = 0.5^\circ$, $\theta_3 = 0.35^\circ$, $\theta_4 = 0.24^\circ$, and $\theta_5 = 0.2^\circ$.  At these angles, $v_F(\theta)$ also vanishes.  Another model is presented in Ref.~\cite{Tarnopolsky_19}; this chirally symmetric model, which we will refer to as the TKV model, assumes $w_0 = 0$ and $w_1 = 110$~meV, and develops a fundamental continuum model for TBG which features not just the vanishing of the Fermi velocity, but also the perfect flattening of the entire lowest band.  When parametrized in terms of $\alpha$, the magic angles recur with a periodicity of $\Delta \alpha \approx 3/2$.  It also develops a perturbation formula for the renormalized Fermi velocity:
\begin{equation}   \label{eq:fermi_velocity-Tarnopolsky}
\frac{v_F(\theta)}{v_0} =
\frac{1 - 3 \alpha^2(\theta) + \alpha^4(\theta) - \frac{111}{49} \, \alpha^6(\theta) + \frac{143}{294} \, \alpha^8(\theta)}
  {1 + 3 \alpha^2(\theta) + 2 \alpha^4(\theta) + \frac{6}{7} \, \alpha^6(\theta) + \frac{107}{98} \, \alpha^8(\theta)} .
\end{equation}
The first magic angle in Ref.~\cite{Tarnopolsky_19} is $\theta_1 = 1.09^\circ$ [$\alpha(\theta_1) \approx 0.586$], and the second magic angle is $\theta_2 \approx 0.29^\circ$ [$\alpha(\theta_2) \approx 2.12$].  Although the first magic angle calculated in Refs.~\cite{bistritzer2011moire} and \cite{Tarnopolsky_19} is similar, the second magic angle of \cite{Tarnopolsky_19} is substantially different than that of Ref.~\cite{bistritzer2011moire}.  The End Matter (EM) contains a derivation of the renormalized Fermi velocity with $w_0 \ne w_1$.  Reference~\cite{Bennett_24} applied first-principles density functional theory to the TBG model of Ref.~\cite{bistritzer2011moire} and demonstrated that the renormalized Fermi velocity $v_F(\theta)$ remains finite even at the magic angles, where the lower band becomes extremely flat and the Fermi velocity attains a nonzero minimum value, $v_F(\theta_1) \approx 0.006 \, v_0$.  Reference~\cite{Carr_19} applied an ab initio ${\mathbf k} \cdot {\mathbf p}$ perturbation continuum model for TBG which accurately accounts for the effects of atomic relaxation.  The authors demonstrated that the Fermi velocity does not vanish within the range of the second and third magic angles.

\begin{figure}
\includegraphics[width=0.9 \linewidth,angle=0] {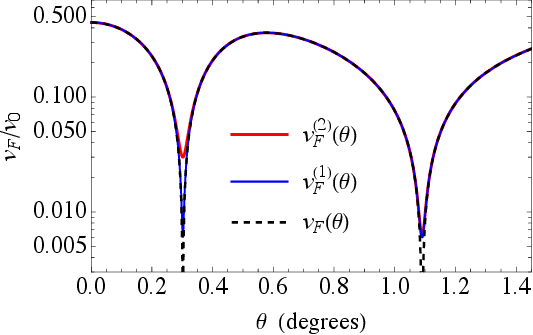}
\caption{\footnotesize
Renormalized Fermi velocity $v_F(\theta)$ versus $\theta$ using the TKV model \cite{Tarnopolsky_19} (dased black curve), which vanishes at the magic angles, $\theta_1$ and $\theta_2$.  The solid blue and red curves show $v_F(\theta)$ which remain finite at $\theta_1$ and $\theta_2$, as in Ref.~\cite{Bennett_24} (which considers only the first magic angle).  The blue curve has the same Fermi velocity at $\theta_1$ and $\theta_2$, $v_F(\theta_1) = v_F(\theta_2) = 0.006 \, v_0$, and the red curve has $v_F(\theta_1) = 0.006 \, v_0$, and $v_F(\theta_2) = 0.03 \, v_0$.
}
\label{FigvFTKV}
\end{figure}

Figure \ref{FigvFTKV} shows the renormalized Fermi velocity $v_F(\theta)$ versus $\theta$ given by Eq.~(\ref{eq:fermi_velocity-Tarnopolsky}) (the black dashed curve).  $v_F(\theta)$ vanishes at the first and second magic angles, $\theta_1 = 1.09^\circ$ and $\theta_2 = 0.302^\circ$ \cite{Tarnopolsky_19}.  The solid blue and red curves were calculated using a model with renormalized Fermi velocities that do not vanish at the magic angles~\cite{Bennett_24},
\begin{equation}   \label{eq:tilde-v_F}
  \tilde{v}_F(\theta) = \sqrt{v_F^2(\theta) + v_{\mathrm{min}}^2(\theta)} ,
\end{equation}
where $v_{\mathrm{min}}(\theta) = v_1 \Theta(\delta\theta_1 - |\theta - \theta_1|) + v_2 \Theta(\delta\theta_2 - |\theta - \theta_2|)$, $\delta\theta_1 = 0.02^\circ$, $\delta\theta_2 = 0.04^\circ$~\cite{delta-theta}, and $\Theta(\bullet)$ is the Heaviside step function.  Note that $v_2$ is not well established in the literature. The blue curve assumes that $v_1 = v_2 = 0.006 \, v_0$ \cite{Bennett_24}, and the red curve assumes that $v_1 = 0.006 \, v_0$, and arbitrarily assumes that $v_2 = 0.03 \, v_0$ (see Fig.~5 of Ref.~\cite{Carr_19}).

{\it Berry connection and the AC phase}---The effective low-energy Hamiltonian in Eq.~(\ref{eq:h_eff-def}) can be rewritten as
\begin{equation}   \label{eq:h_eff-Berry-connection}
h_{\text{eff}}(\hbar \mathbf{k}) =
\hbar \tilde{v}_F(\theta) \, \bm{\tau} \cdot \big[ \sigma_0  \bm{k} - {\bf A}(\theta) \big] ,
\end{equation}
where the SU(2) Berry connection is,
%
\begin{equation} \label{defA_theta}
{\bf A}(\theta) = \frac{\lambda_R}{\hbar \tilde{v}_F(\theta)} \, \hat {\bf z} \times {\bm \sigma} ,
\end{equation}
and $\tilde{v}_F(\theta)$ is the renormalized Fermi velocity in Eq.~(\ref{eq:tilde-v_F}) which does not vanish at the magic angles.

\begin{figure}
\includegraphics[width=0.9 \linewidth,angle=0] {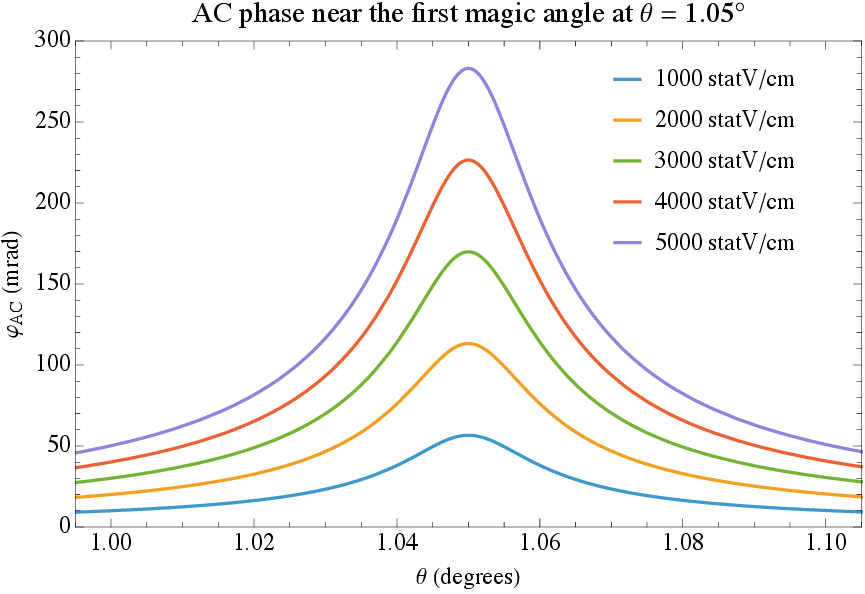}
\caption{\footnotesize
AC phase $\varphi_{\mathrm{AC}}(\theta)$ near the first magic angle $\theta \approx 1.05^\circ$ using the Bistritzer-MacDonald model~\cite{bistritzer2011moire} with corrected Fermi velocity at the magic angle \cite{Bennett_24}, $\tilde{v}_F(\theta_1) = 0.006 \, v_0$, for five values of electric field strength.}
\label{Fig_ACphase_theta=1.05}
\end{figure}

The wave function of an electron propagating from a point $\bm{r}_i$ to a point $\bm{r}_f$, acquires an Aharonov-Casher (AC) phase factor ${\mathcal U}_{\mathrm{AC}}$ given by
\begin{equation}   \label{eq:AC-phase-def}
{\mathcal U}_{\mathrm{AC}} =
{\mathcal P}{\rm exp} \Big( - i \int_{\mathbf{r}_i}^{\mathbf{r}_f} {\bf A}(\theta) \cdot d\mathbf{r} \Big) ,
\end{equation}
where we take the integration path to be a straight line connecting $\mathbf{r}_i$ and $\mathbf{r}_f$.  Since ${\bf A}(\theta)$ in Eq.~(\ref{eq:AC-phase-def}) is independent of $\mathbf{r}$, it can be removed from the integral, hence the AC phase factor is Abelian and is given by \cite{non-Abelian-phase-factor}
\begin{equation}   \label{eq:AC-phase-res}
{\mathcal U}_{\mathrm{AC}} =
\exp
\Big(
  - \frac{i \lambda_R}{\hbar \tilde{v}_F(\theta)} \, \mathbf{L} \cdot {\bm \sigma}
\Big) ,
\end{equation}
where $\mathbf{L} = ( \mathbf{r}_f - \mathbf{r}_i) \times \hat {\bf z}$.  If the electron spin wave function is an eigenfunction of $\hat{\mathbf{L}} \cdot {\bm \sigma}$ with eigenvalue of $\sigma = \pm 1$, where $\hat{\mathbf{L}} = \mathbf{L}/L$ is the unit vector in the direction of $\mathbf{L}$, and $L = |\mathbf{L}|$ is the path-length, the AC phase is given by $\varphi_\sigma(\theta) = \sigma \varphi_{\mathrm{AC}}(\theta)$, where
\begin{equation}   \label{eq:AC-phase}
\varphi_{\mathrm{AC}}(\theta) = \frac{\lambda_R L}{\hbar \tilde{v}_F(\theta)} .
\end{equation}
Figure~\ref{Fig_ACphase_theta=1.05} plots the AC phase $\varphi_{\mathrm{AC}}(\theta)$ versus twist angle $\theta$ for $L = 0.1 \, \mu$m and renormalized Fermi velocity $\tilde{v}_F(\theta)$ given by the BMD model in Eq.~(\ref{eq:fermi_velocity-Bistritzer}), with corrected value of the Fermi velocity at the magic angle, $\tilde{v}_F(\theta_1) = 0.006 \, v_0$ \cite{Bennett_24}.  The AC phase is a linear function of the electric field strength, $E$, and exhibits a sharp peak at the magic angle, $\theta_1 = 1.05^\circ$.

\begin{figure}
\includegraphics[width=0.95 \linewidth,angle=0] {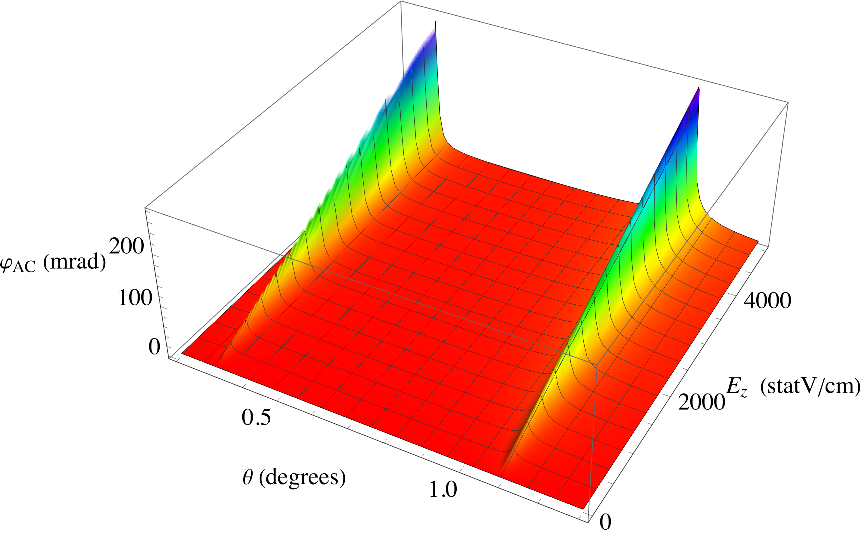}
\caption{\footnotesize
AC phase near the first two magic angles $\theta \approx 1.08^\circ, 0.3^\circ$ versus $\theta$ and electric field strength $E_z$ using the TKV model \cite{Tarnopolsky_19} with corrected Fermi velocity at the magic angles \cite{Bennett_24}, $\tilde{v}_F(\theta_1) = \tilde{v}_F(\theta_2) = 0.006 \, v_0$ (where the minimal Fermi velocity for the two magic angles has been assumed to be equal).}
\label{FigACphase2MagicAngles}
\end{figure}

Figure~\ref{FigACphase2MagicAngles} plots the AC phase, $\varphi_{\mathrm{AC}}(\theta)$, versus the twist angle $\theta$ and the electric field strength $E$ for $L = 0.1 \, \mu$m and the renormalized Fermi velocity $\tilde{v}_F(\theta)$ which is given by the TKV model in Eq.~(\ref{eq:fermi_velocity-Tarnopolsky}), with corrected values at the magic angles $\theta_1 = 1.09^\circ$ and $\theta_2 = 0.302^\circ$.  Since the renormalized Fermi velocity at $\theta_2$ is unavailable in the literature, in Fig.~\ref{FigACphase2MagicAngles} we have taken $\tilde{v}_F(\theta_2) = \tilde{v}_F(\theta_1) = 0.006 \, v_0$~(\cite{Bennett_24}.  The AC phase is a linear function of the electric field strength, $E$, and exhibits a sharp peak at the magic angles.  The height of these peaks is inversely proportional to the minimum of the Fermi velocity $\tilde{v}_F(\theta_1) [ = \tilde{v}_F(\theta_2)]$.  Figure~\ref{FigACphase2MagicAngles} shows the same peak of $\varphi_{\mathrm{AC}}(\theta)$ at $\theta = \theta_1$ and $\theta_2$.  The EM contains a similar figure with $\tilde{v}_F(\theta_1) = 0.006 \, v_0$ and $\tilde{v}_F(\theta_2) = 0.03 \, v_0$, hence the AC phase peak at $\theta_2$ is a fifth of the peak at $\theta_1$.

{\it Double-tip scanning tunneling spectroscopy (DTSTS)}--DTSTS is an advanced surface analysis technique that utilizes two independently controlled tips used to measure quantum electron correlations, and directly evaluate the single-electron Green's function of surfaces \cite{Dana-2008, Ji-2012, Settnes-2014, Su-2025}; by applying different bias voltages to each tip, researchers can track transport properties across the nanometer scale.  Here we propose using DTSTS with ferromagnetic tips to measure the AC phase in TBG.  The source tip is placed at position $\mathbf{r}_s$ and biased with voltage of $V_s$, and the drain tip is placed at position $\mathbf{r}_d$ and biased with voltage of $V_d$.  Taking $\mathbf{r}_d - \mathbf{r}_s = L \, \hat{\mathbf{x}}$, so that $\mathbf{L} = L \, \hat{\mathbf{y}}$, the AC phase factor in Eq.~(\ref{eq:AC-phase-res}) commutes with the spin matrix $\sigma_y$.  Taking the source tip polarized along the $z$-axis, the spin wave function of the electron that tunnels from the source tip to the TBG is also polarized in the $z$-direction,
\begin{equation}   \label{eq:wf-s}
| \psi(\mathbf{r}_i) \rangle =
| \chi_s \rangle =
\frac{i}{\sqrt{2}} \,
\big\{
  | \chi^{(y)}_{\uparrow} \rangle -
  | \chi^{(y)}_{\downarrow} \rangle
\big\} ,
\end{equation}
where $| \chi_s \rangle$ is the eigenfunction of $\sigma_z$ with eigenvalue $\sigma = 1$, and
$| \chi^{(y)}_{\uparrow} \rangle$ and $| \chi^{(y)}_{\downarrow} \rangle$ are eigenfunctions of $\sigma_y$ with eigenvalues $\sigma = \pm 1$, respectively.  At the drain tip position, $\mathbf{r}_d$, the spin wave function $|\chi^{(y)}_{\uparrow}\rangle$ acquires the AC phase $\varphi_{\mathrm{AC}}(\theta)$ which is given in Eq.~(\ref{eq:AC-phase}), and the spin wave function $|\chi^{(y)}_{\downarrow}\rangle$ acquires the AC phase $-\varphi_{\mathrm{AC}}(\theta)$.  Hence the spin wave function at the drain position is
\begin{equation}   \label{eq:wf-d}
| \psi(\mathbf{r}_d) \rangle = \frac{i}{\sqrt{2}} \,
\big[
  e^{\varphi_{\mathrm{AC}}(\theta)} \, | \chi^{(y)}_{\uparrow} \rangle -
  e^{-\varphi_{\mathrm{AC}}(\theta)} | \chi^{(y)}_{\downarrow} \rangle
\big] .
\end{equation}
Taking the drain tip to be polarized along the $x$-axis, the probability that an electron tunnels into the drain tip is,
\begin{equation}   \label{eq:P-res}
P(\theta) =
\big| \langle \chi_d | \psi(\mathbf{r}_d) \rangle |^2 =
\frac{1 + \sin[2 \varphi_{\mathrm{AC}}(\theta)]}{2} ,
\end{equation}
where $|\chi_d \rangle$ is an eigenfunction of $\sigma_x$ with eigenvalue $\sigma = 1$, $|\chi_d \rangle =
\frac{1}{\sqrt{2}} \, \big\{ | \chi^{(y)}_{\uparrow} \rangle + | \chi^{(y)}_{\downarrow} \rangle \big\}$.  For small AC phase, $\varphi_{\mathrm{AC}}(\theta) \ll 1$, $P(\theta) \approx 1/2 + \varphi_{\mathrm{AC}}(\theta)$, i.e., the probability $P(\theta)$ is a linear function of the AC phase, $\varphi_{\mathrm{AC}}(\theta)$.  Therefore, measuring the probability $P(\theta)$ yields an effective measurement of the AC phase.

{\it Conclusions}---Measuring the AC phase in TBG, particularly near magic angles, where the bands are nearly flat, can help us understand the strong electron correlations, nontrivial topology, and tunable spin-orbit interactions.  This offers several advantages in condensed matter physics and quantum device engineering.  The AC phase in TBG is inversely proportional to the renormalized Fermi velocity. At the magic angles, the Fermi velocity  becomes small, therefore the AC phase becomes large.  By measuring the AC phase, researchers can directly map the valley Berry curvature and quantum metric of the flat bands \cite{Peotta_15}.  This provides a quantitative probe of valley-contrasting topology.  Quantum interference loops that are sensitive to the AC phase can detect the nontrivial winding numbers and fractional statistics of these spin-textured quasiparticles as they navigate the moir\'e superlattice.  Because the AC phase depends explicitly on local electric field gradients $\boldsymbol{\nabla} E$, local phase shifts reflect variations in electron density and built-in potentials across individual moir\'e AA/AB stacking regions.  Observing local AC phase variations (e.g., via scanning tunneling interferometry or SQUID-on-tip techniques) acts as a microscopic diagnostic tool for strain-induced pseudo-magnetic fields and moir\'e lattice relaxation.


\newpage
\begin{widetext}
\begin{center}
\textbf{End Matter}
\end{center}
\end{widetext}

We expand the discussion of the main text (MT) and address the following topics: (1) derivation of the effective low-energy Hamiltonian of a twisted bilayer graphene (TBG) and calculation of the effective Fermi velocity $v_F(\theta)$ which depends on the twist angle $\theta$,  and (2) the Aharonov-Casher (AC) phase $\varphi(\theta)$ for $\theta$ close to the first and second magic angles, $\theta_1$ and $\theta_2$.

{\it Renormalized Fermi velocity}---We apply second-order degenerate perturbation theory in the moir\'e potential to derive an effective low-energy Hamiltonian near the Dirac point of layer 1 ($\mathbf{k} = \frac{\bm{q}_1}{2}$), setting the origin at this Dirac point, and a renormalized Fermi velocity.  At $\mathbf{k} = 0$ (relative to layer 1's Dirac point), the interlayer tunneling $T(\bm{r})$ couples the state in layer 1 to three degenerate states in layer 2 that are shifted by the interlayer momentum transfer vectors $\mathbf{q}$, see Eq.~(\ref{eq:q_1-q_2-q_3}).  The following Dirac equation for low-energy states is derived using second-order degenerate perturbation theory with respect to the moir\'e potential $T(\bm{r})$,
\begin{equation}   \label{eq:Dirac-low-energy}
h_{\text{eff}}(\hbar \mathbf{k}) \psi(\mathbf{k}) = \epsilon \, \psi(\mathbf{k}) ,
\end{equation}
where  $h_{\text{eff}}(\hbar \mathbf{k})$ is the effective Hamiltonian modified by a self-energy term, denoted by $\Sigma(\epsilon, \mathbf{k})$:
\begin{equation}  \label{eq:H_eff-self-energy}
h_{\text{eff}}(\hbar \mathbf{k}) = h(\hbar\mathbf{k}) + \Sigma(\epsilon, \mathbf{k}) ,
\end{equation}
and where $h(\hbar {\mathbf k})$ is the kinetic energy Hamiltonian in the momentum representation, see Eq.~(3), and the self-energy due to tunneling into layer 2 is given by
\begin{equation}   \label{eq:self-energy}
\Sigma(\epsilon, \mathbf{k}) =
\sum_{j=1}^3 T_j \frac{1}{\epsilon - h(\hbar \mathbf{k} - \hbar \mathbf{q}_j)} T_j^\dagger .
\end{equation}
Let us consider the low energy states, for which $|\epsilon| \ll \hbar v_0 k_\theta$, and $|\mathbf{k}| \ll k_\theta$, where $k_\theta$ is defined in the first equation of Eq.~(\ref{eq:k_theta}).  Expanding the self-energy in $\epsilon$ and $\mathbf{k}$, and keeping the linear corrections, we obtain
\begin{align}
&
\Sigma(\epsilon, \mathbf{k}) \approx
- \frac{1}{2 \hbar v_0 k_\theta^2}
\sum_{j=1}^3 T_j \, (\bm{q}_j \cdot \boldsymbol{\tau}) \, T_j^\dagger
\nonumber \\ & \quad
- \frac{\epsilon}{\hbar^2 v_0^2 k_\theta^2} \sum_{j=1}^3 T_j T_j^\dagger
- \frac{1}{\hbar v_0 k_\theta^2}
\sum_{j=1}^3 T_j (\bm{k} \cdot \boldsymbol{\tau}) T_j^\dagger .
\end{align}
Due to the $C_3$ rotational symmetry of the TBG, the first term on the right-hand side of this equation vanishes when summed over $j$, while the second and third terms are simplified as follows:
\begin{align}   \label{eq:Sigma_epsilon-res}
\Sigma(\epsilon, \mathbf{k}) &\approx
- \frac{3 (w_0^2 + w_1^2)}{\hbar^2 v_0^2 k_\theta^2} \, \epsilon
- \frac{3 w_1^2}{\hbar v_0 k_\theta^2} \, (\bm{k} \cdot \boldsymbol{\tau}) .
\end{align}
Using Eqs.~(\ref{eq:H_eff-self-energy}), (\ref{eq:Sigma_epsilon-res}), Eq.~(\ref{eq:Dirac-low-energy}) takes the form
\begin{equation}   \label{eq:Dirac-alpha}
\big[ 1 - 3 \alpha_1^2(\theta) \big] \, \sigma_0 \, \big( \bm{\tau} \cdot \bm{k} \big) \, \psi(\mathbf{k}) =
\big[ 1 + 3 \alpha_1^2(\theta) + 3 \alpha_0^2(\theta) \big] \epsilon \, \psi(\mathbf{k}) ,
\end{equation}
where
\begin{equation}   \label{eq:alpha_l}
\alpha_l(\theta) =
\frac{w_l}{\hbar v_0 k_\theta} =
\frac{w_l}{2\hbar v_0 k_D \sin(\theta/2)} ,
\end{equation}
and $l = 0, 1$.  Dividing the left- and right-hand sides of Eq.~(\ref{eq:Dirac-alpha}) by $1 + 3 \alpha_1^2(\theta) + 3 \alpha_0^2(\theta)$, we obtain the following Dirac equation,
\begin{equation}
\tilde{h}_{\text{eff}}(\hbar \mathbf{k}) \, \psi(\mathbf{k}) = \epsilon \, \psi(\mathbf{k}) ,
\end{equation}
where the effective Hamiltonian $\tilde{h}_{\text{eff}}(\mathbf{k})$ is given by,
\begin{equation}
\tilde{h}_{\text{eff}}(\hbar \mathbf{k}) = \hbar v_F(\theta) \, \sigma_0 \, \bm{\tau} \cdot \bm{k} ,
\end{equation}
and the renormalized Fermi velocity is~\cite{lopes2007graphene, bistritzer2011moire, Tarnopolsky_19}
\begin{equation}   \label{eq:v_F-theta-w_0-w_1}
v_F(\theta) = v_0 \,
\frac{1 - 3 \alpha_1^2(\theta)}
{\displaystyle 1 + 3 \alpha_0^2(\theta) + 3 \alpha_1^2(\theta)} .
\end{equation}
Setting $w_0 = w_1 \equiv w$, Eq.~(\ref{eq:v_F-theta-w_0-w_1}) reduces to Eq.~(12) in the MT, which was first published  in Ref.~\cite{bistritzer2011moire}.  Setting $w_0 = 0$, Eq.~(\ref{eq:v_F-theta-w_0-w_1}) provides an approximation to Eq.~(14) in the MT as obtained in Ref.~\cite{Tarnopolsky_19}.

\begin{figure}
\includegraphics[width=0.95 \linewidth,angle=0] {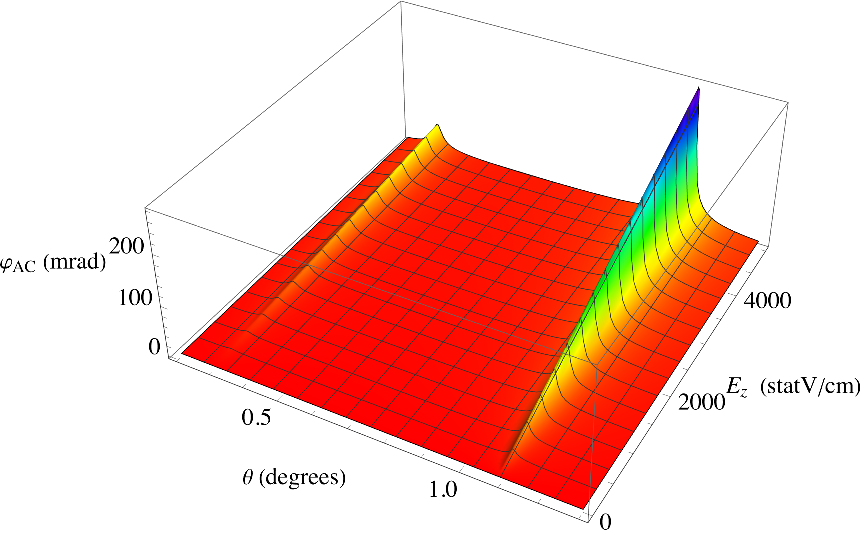}
\caption{\footnotesize
AC phase near the first two magic angles $\theta \approx 1.08^\circ, 0.3^\circ$, as in Fig.~3, but with the minimal Fermi velocity at the second magic angle taken to be five times larger than the value at the first magic angle.}
\label{FigACphase2MagicAngles2vF}
\end{figure}

{\it AC phase}---In contrast with Fig. 3, Fig.~\ref{FigACphase2MagicAngles2vF} shows the AC phase $\varphi_{\mathrm{AC}}(\theta)$ versus $\theta$ and $E$ for $\tilde{v}_F(\theta_1) = 0.006 \, v_0$ \cite{Bennett_24} and $\tilde{v}_F(\theta_2) = 0.03 \, v_0$.  Since $\tilde{v}_F(\theta_2)$ is five times larger than $\tilde{v}_F(\theta_1)$, the AC phase $\varphi_{\mathrm{AC}}(\theta_2)$ in Fig.~\ref{FigACphase2MagicAngles2vF} is smaller than $\varphi_{\mathrm{AC}}(\theta_1)$ by a factor of five.

\clearpage

\clearpage


\begin{thebibliography}{99}
\bibitem{delta-theta}
The parameters $\delta\theta_l$ with $l = 1, 2$ are such that $|v_F(\theta_l \pm \delta\theta_l)| \gg v_{\min}(\theta_l)$, so that $\tilde{v}_F(\theta_l \pm \delta\theta_l) \approx |v_F(\theta_l \pm \delta\theta_l)|$.
\bibitem{non-Abelian-phase-factor}
In general, the Berry connection, $\mathbf{A}(\mathbf{r}, \theta)$, is a function of $\mathbf{r}$, and the path is a curve, hence the AC phase factor is non-Abelian \cite{paper-AC-PRB-short, paper-AC-PRB-long}.  In this paper, however, $\mathbf{A}(\theta)$ is independent of $\mathbf{r}$, and the path is a straight line, therefore the AC phase factor is Abelian.
\end{thebibliography}
\end{document}